\documentclass[%
 reprint,
 amsmath,amssymb,
 aps,
prb,
]{revtex4-2}

\usepackage{graphicx}% Include figure files
\usepackage{dcolumn}% Align table columns on decimal point
\usepackage{bm}% bold math
\usepackage[mathlines]{lineno}% Enable numbering of text and display math
\begin{document}

%\preprint{APS/123-QED}

\title{Dependence of strength of spin- orbit interaction on polarity of interface}% Force line breaks with \\

%\title{Systematic study of modulation of Spin- orbit interaction by gate voltage in FeCoB nanomagnets}% Force line breaks with \\

\author{Vadym Zayets}
\affiliation{National Institute of Advanced Industrial Science and Technology (AIST), Umezono 1-1-1, Tsukuba, Ibaraki, Japan}

\date{\today}% It is always \today, today,

\begin{abstract}
It was experimentally observed that both magnetic anisotropy and spin-orbit interaction strength change when the magnetization of the nanomagnet is reversed. This indicates a variation in spin-orbit interaction strength depending on whether the magnetic field penetrates the interface from a ferromagnetic to a non-magnetic metal or vice versa. Systematic measurements of over 100 nanomagnets revealed a consistent, yet unexpected, pattern between variations in magnetic anisotropy and spin-orbit interactions with magnetization reversal. These changes align along a single straight line with a negative slope, suggesting a complex and indirect relationship. Our findings also suggest the presence of an additional, yet-to-be-identified effect that influences the change in magnetic anisotropy with magnetization reversal, beyond the variations in spin-orbit interaction strength. This finding highlights the complexity of magnetic behavior at the nanoscale and the critical role of magnetization direction in determining anisotropic properties.
\end{abstract}

 %Use showkeys class option if keyword

\maketitle

%\linenumbers\relax % Commence numbering lines

The spin-orbit interaction is one of the most fundamental effects in physics, playing a pivotal role in numerous phenomena within solid state physics.In the realm of magnetism, the significance of the spin-orbit interaction is of utmost importance, as it gives rise to magnetic anisotropy, which is essential for the existence of permanent magnets. All type of magnetic memory relies on magnetic anisotropy, as it creates two stable magnetization directions in a nanomagnet, allowing a bit of data to be stored based on these two magnetization states \cite{Johnson1996Hani,MRAM2020Everspin,MRAM2017Ohno,MRAMCubukcu}.

Understanding the features, properties, and particularities of spin-orbit interaction is essential for both fundamental physics and data storage technologies. In this work we study one such particularity: the dependence of the strength of the spin-orbit interaction on the polarity of the interface. Consequently, we examine the reason why and how the strength of the spin-orbit interaction and the magnetic anisotropy change with the reversal of the magnetization direction.

\begin{figure}[tb]
	\begin{center}
		\includegraphics[width=7 cm]{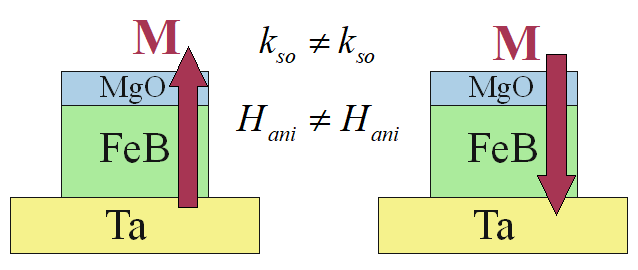}
	\end{center}
	\caption{
		{Two stable magnetization directions of an FeB nanomagnet due to its interfacial magnetic anisotropy. The strengths of the spin-orbit interaction ($k_{so}$) and the anisotropy field ($H_{ani}$) vary depending on whether the magnetization is directed from the ferromagnetic material towards the non-magnetic material or in the opposite direction. Given that the FeB/Ta and FeB/MgO interfaces contribute differently to the anisotropy and spin-orbit interaction, the overall values of $k_{so}$ and $H_{ani}$ in Ta/FeB/MgO nanomagnets differ for opposite magnetization directions.} 
	}
	\label{fig:MupMdown} 
\end{figure}

Figure \ref{fig:MupMdown} shows the cross-section of a Ta/FeB/MgO nanomagnet in its two stable magnetization directions (up and down). Our experimental observations reveal that the measured strength of spin-orbit interaction ($k_{so}$) and the anisotropy field ($H_{ani}$) differ for these two magnetization directions. Given that the magnetic anisotropy in FeB nanomagnets is primarily of interfacial origin, this implies that the strength of the spin-orbit interaction depends on the direction in which the magnetic field traverses the interface. Specifically, for the FeB/Te interface, the strength of the spin-orbit interaction differs when the magnetic field is directed from Ta to FeB compared to when it is directed from FeB to Ta.

%%%%%%%%%%%%%%%%%%%%%%%%%%%%%%%%%%%%
%%%%%%%%%%%%%%%%%%%%%%%%%%%%%%%%%%%%%%%%%%%%%%%%%%%%%%%%%%%%%%%%%%%%%

A similar difference exists at the FeB/MgO interface, where both the strength of the spin-orbit interaction and the interfacial magnetic anisotropy differ for opposite magnetization directions. 

In a symmetrical nanomagnet (e.g., Ta/FeB/Ta), the contributions from the lower and upper interfaces to the polarity dependence of the spin-orbit interaction strength are opposite, resulting in no net polarity dependence. However, in the studied asymmetrical nanomagnet (Ta/FeB/MgO), the contributions from the Ta/FeB and FeB/MgO interfaces differ and do not cancel each other out, leading to a polarity-dependent overall $k_{so}$ and $H_{ani}$.

All measurements were conducted using a recently introduced novel method for assessing the strength of the spin-orbit interaction \cite{Zayets2024SObasic}. Experimental measurements of the modulation of spin-orbit interaction strength by an electrical current \cite{ZayetsArch2024_SO_SOT} and a gate voltage \cite{ZayetsArch2024_SO_VCMA} have been reported. This method has proven to be robust, reliable, and repeatable, offering valuable insights into the complex fundamental phenomena of spin-orbit interaction and magnetic anisotropy. This measurement technique is based on a measurement of the strength of magnetic anisotropy under an external magnetic field applied along the magnetic easy axis \cite{Zayets2024SObasic}. Since the strength of magnetic anisotropy \cite{Johnson1996Hani, LandauField} is inherently linked to the strength of spin-orbit interaction (SO), measuring SO strength through measurements of magnetic anisotropy is  the most intuitive and direct approach for measuring SO strength. The parameter that describes the strength of magnetic anisotropy is the anisotropy field $H_{ani}$. Both theoretical and experimental evidence  \cite{Zayets2024SObasic} indicate that the anisotropy field increases linearly with the external magnetic $H_z$ field applied along the magnetic easy axis. This relationship can be expressed as:

\begin{equation}
	H_{ani}-H_z= H^0_{ani}+k_{so}H_z
	\label{EqHaniHz}
\end{equation} 
where $k_{so}$ is the coefficient of spin- orbit interaction, which defines the strength of spin-orbit interaction, and $H^0_{ani}$ is the anisotropy field in absence of $H_z$.

%%%%%%%%%%%%%%%%%%%%%%

%%%%%%%%%%%%%%%%%%%%%%%%%%%%%%%%%%%%
\begin{figure}[tb]
	\begin{center}
		\includegraphics[width=6.5 cm]{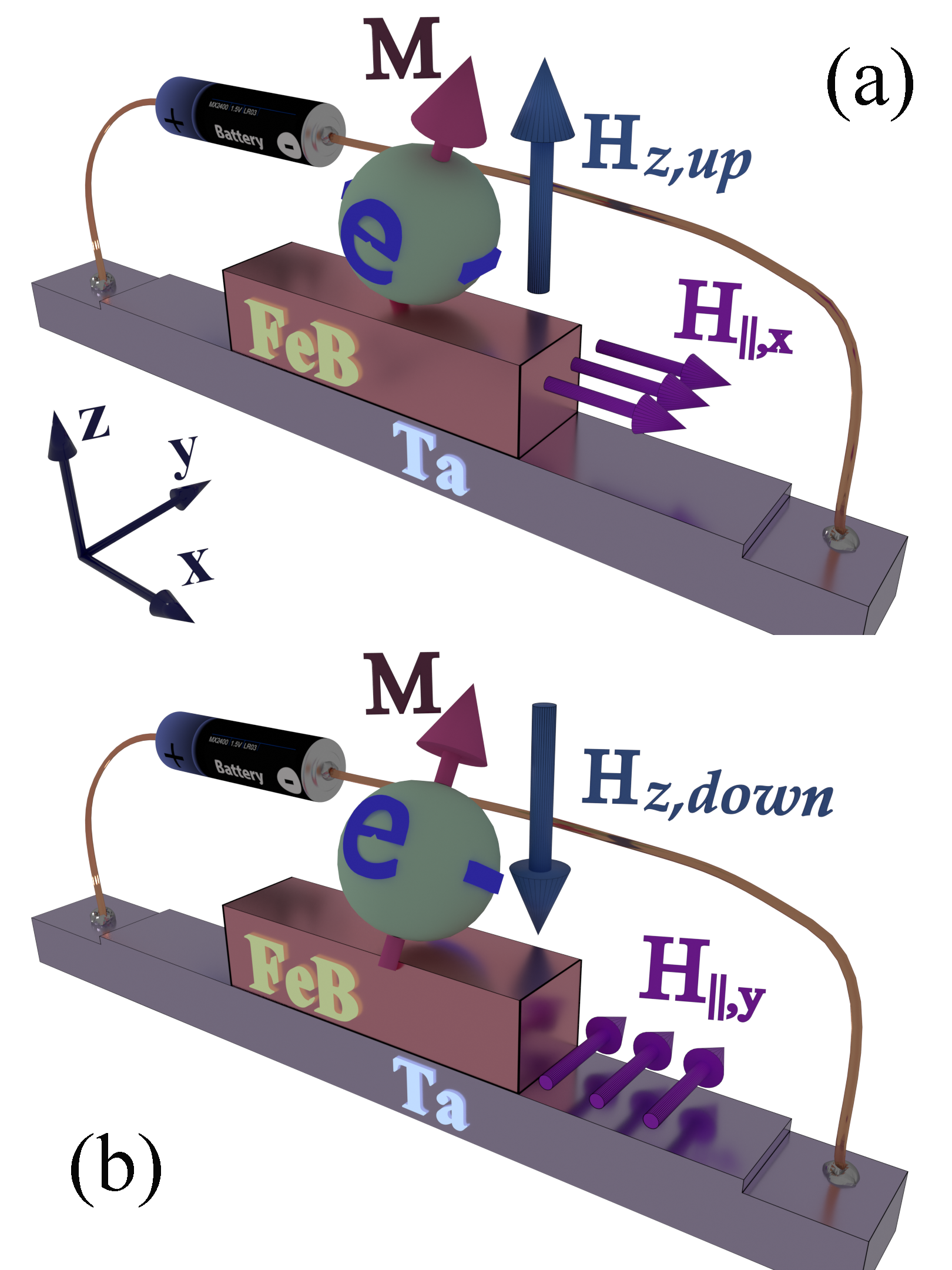}
	\end{center}
	\caption{
		{Experimental setup for measuring the strength of spin-orbit interaction and anisotropy field. (a) The bias magnetic field $H_z$ is directed upwards, while the in-plane magnetic field $H_{||,x}$ is scanned along the direction of the electrical current. (b) The bias magnetic field $H_z$  is directed downwards, while the in-plane magnetic field $H_{||,y}$ is scanned perpendicular to the current.} 
	}
	\label{fig:ExpVertical} 
\end{figure}
%%%%%%%%%%%%%%%%%%%%%%

To investigate the dependence of spin-orbit interaction strength on interface polarity, a FeB nanomagnet was fabricated atop a Ta nanowire with an attached Hall probe aligned to the nanomagnet. The nanomagnet was covered by a MgO layer. Nanomagnets of varying sizes, ranging from 50 nm x 50 nm to 2000 nm x 2000 nm, were fabricated at different locations on a single wafer. The equilibrium magnetization of all nanomagnets was perpendicular- to- plane. The equilibrium magnetization of all nanomagnets was perpendicular- to- plane. A low current density of  4 $mA/\mu ^2$ was used to prevent any influence from spin accumulation \cite{ZayetsArch2024_SO_SOT, Zayets2020MishenkoSpinPol,ZayetsJMMM2018Holes,ZayetsJMMM2023Parametric}.

The experiments were conducted at room temperature, well below the Curie temperature of FeB. The measurement procedure involved recording the Hall angle while sweeping an in-plane external magnetic field $H_x$ (see Fig. \ref{fig:ExpVertical}). A perpendicular-to-plane magnetic field $H_z$ was used as a parameter and kept constant during the $H_x$ sweep. The magnetic field $H_x$, which is applied perpendicularly to the magnetic easy axis,  tilts the magnetization of the nanomagnet. The tilt angle was determined by measuring the reduction in Hall voltage.  The anisotropy field $H_{ani}$ was evaluated by fitting the linear relationship between the measured in-plane magnetization component $M_x$ and $H_x$. As expected, $H_{ani}$ increased with the external field $H_z$. The coefficient of spin-orbit interaction $k_{so}$ and the anisotropy field $H^0_{ani}$ in the absence of an external magnetic field were evaluated by fitting Eq. \ref{EqHaniHz} to the measurement data of $H_{ani}$ vs. $H_z$.  All details of the measurement procedure are described in Ref. \cite{Zayets2024SObasic}.

%%%%%%%%%%%%%%%%%%%%%%%%%%%%%%%%%%%%
\begin{figure}[tb]
	\begin{center}
		\includegraphics[width=7 cm]{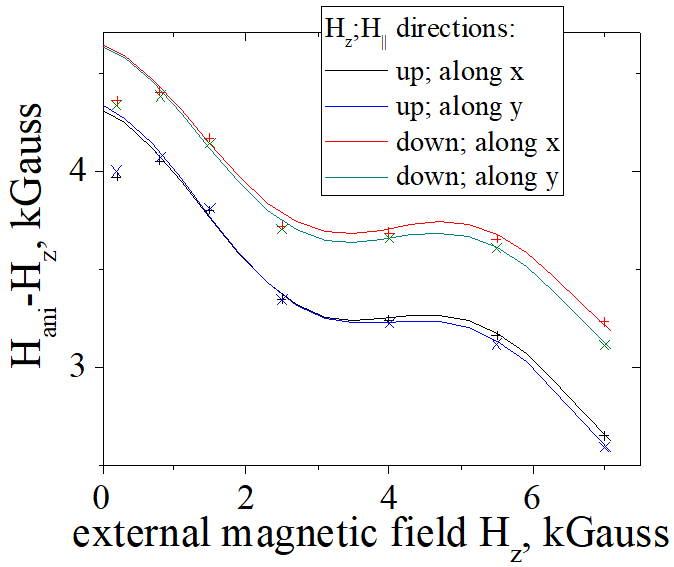}
	\end{center}
	\caption{
		{Anisotropy field $H_{ani}$ as a function of external perpendicular-to-plane magnetic field $H_{z}$ measured for two opposite directions of the bias magnetic field $H_{z}$ and two opposite directions of the scanned in-plane magnetic field $H_{||}$. Sample: $Ta(2 nm)/ FeB(1.3 nm)/ MgO(5.1 nm)$.} 
	}
	\label{fig:FigHaniHz} 
\end{figure}
%%%%%%%%%%%%%%%%%%%%%%

For each nanomagnet, four measurements were conducted. The in-plane magnetic field was scanned either along or perpendicular to the nanowire, while the perpendicular-to-plane magnetic field was directed either upward or downward. If the strength of the spin-orbit interaction were independent of the interface polarity, the results of all four measurements should be identical. Figure \ref{fig:FigHaniHz} shows the results of the measurements of $H_{ani}$ versus $H_z$. The data is nearly identical for both $H_x$ scans. However, there is a substantial difference in the data for the upward and downward directions of the external magnetic field $H_z$. This clearly indicates a significant dependence on the polarity of the interface with respect to the direction of the magnetic field.

In both cases, the relationship exhibits an approximately linear trend with small oscillations atop the linear dependence. The presence of small oscillations is a well-known characteristic of interfacial spin-orbit interaction \cite{Zayets2024SObasic}. The slope of the line corresponds to the strength of the spin-orbit interaction, described by $k_{so}$. The line offset corresponds to the anisotropy field $H^0_{ani}$ in the absence of an external magnetic field. The offset, and therefore $H^0_{ani}$, is smaller for the upward direction than for the downward direction of $H_z$. In contrast, the slope, and therefore $k_{so}$, is smaller for the upward direction than for the downward direction. The increased gap between the lines at higher $H_z$ clearly indicates an opposite dependence of $k_{so}$ and $H^0_{ani}$ on the polarity of $H_z$.

In all measurements, the applied external magnetic field $H_z$  was significantly greater than the coercive field of the nanomagnets \cite{ZayetsArch2019Hc}. Therefore, the observed differences in the anisotropy field and the strength of the spin-orbit interaction accurately reflect their variations for opposite magnetization directions.

%%%%%%%%%%%%%%%%%%%%%%%%%%%%%%%%%%%%
\begin{figure}[tb]
	\begin{center}
		\includegraphics[width=7 cm]{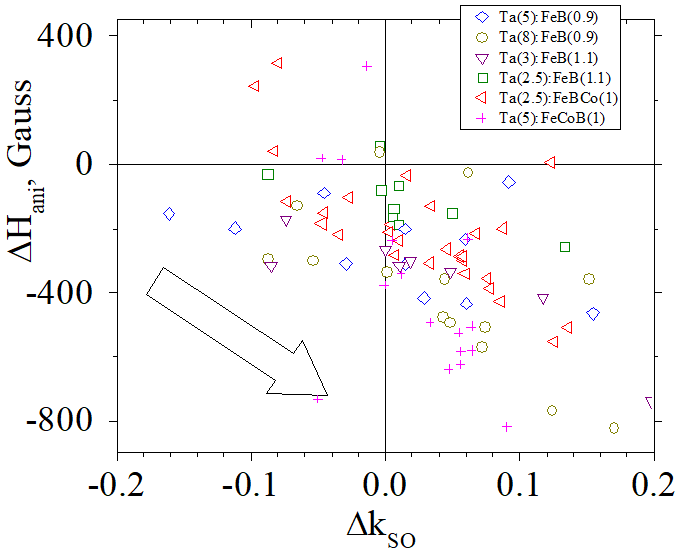}
	\end{center}
	\caption{
		{Difference in the anisotropy field $H^0_{ani}$  between measurements with the external bias magnetic field $H_z$ directed upward and downward, plotted against the corresponding difference in the spin-orbit interaction coefficient $k_{so}$  in nanomagnets of varying sizes and structures. Each dot represents an individual nanomagnet measurement. Dots of the same color and shape correspond to nanomagnets fabricated at different locations on the same wafer. The numbers in parentheses indicate the layer thickness in nanometers.
		} 
	}
	\label{fig:FigDistribution} 
\end{figure}
%%%%%%%%%%%%%%%%%%%%%%

The measured differences in $H^0_{ani}$ and $k_{so}$ for opposite magnetization polarities exhibit a systematic and consistent pattern. In Figure \ref{fig:FigDistribution}, the change in $H^0_{ani}$ for opposite magnetization directions is plotted against the change in $k_{so}$ for nanomagnets of various sizes and structures. The change in $H^0_{ani}$ is mostly negative for all measured nanomagnets, indicating that the anisotropy field is larger when the magnetization is downward. The change in $k_{so}$, however, can be either positive or negative. A notable and intriguing feature of the plot is that all data points align clearly along a straight line with a negative slope, offset from the axis center toward the bottom left. This alignment is observed not only for nanomagnets fabricated on the same wafer but also for those from different wafers.

There is a slight variation in nanomagnet parameters, such as thickness, interface roughness, and sharpness, across a single wafer due to some technological imperfections. This leads to slight variations in the anisotropy field and the strength of the spin-orbit interaction among nanomagnets on the same wafer \cite{Zayets2024SObasic}. It was experimentally observed that improved fabrication technology reduces the spread of the data, but $k_{so}$ and $H^0_{ani}$ still differ substantially for nanomagnets from different wafers. The alignment of data from nanomagnets across different wafers suggests the existence of a universal relationship between $\Delta H_{ani}$ and $\Delta k_{so}$.

Another unexpected feature observed in Fig. \ref{fig:FigDistribution} is the negative slope of the linear trend, given that $H^0_{ani}$ and $k_{so}$  are not independent parameters and are linearly proportional to each other. Therefore, the changes in $H^0_{ani}$ and $k_{so}$  due to magnetization reversal should be proportional, resulting in a positive slope in Fig. \ref{fig:FigDistribution}. The internal magnetic field $H_{int}$  within the nanomagnet correlates  $H^0_{ani}$ and $k_{so}$. This internal magnetic field is responsible for maintaining magnetization along the magnetic easy axis. Given the general similarity between external $H_z$ and internal $H_{int}$ magnetic fields, Eq. \ref{EqHaniHz} can be rewritten as

\begin{equation}
	H_{ani}=(H_z+H_{int})+k_{so} (H_z+H_{int})
	\label{EqHaniHint}
\end{equation} 

where

\begin{equation}
	H^0_{ani}= (1+k_{so})H_{int}
	\label{EqHaniHintkso}
\end{equation} 

Eq. \ref{EqHaniHintkso} indicates a linear proportionality between $H^0_{ani}$  and $k_{so}$. The positive slope of this linear proportionality can be reversed only if, in addition to the spin-orbit interaction, another effect contributes to the change in the anisotropy field due to magnetization reversal. There are known cases \cite{Zayets2024SObasic,ZayetsArch2024_SO_SOT,ZayetsArch2024_SO_VCMA,MMM2022_SO_SOT,Intermag2023_SO_VCMA} where this linear relationship with a positive slope is overridden by another effect.

For example, the modulation of demagnetization due to interface roughness in multi-layer nanomagnets \cite{Zayets2024SObasic} is responsible for the observed negative slope between $H^0_{ani}$  and $k_{so}$. Similarly, the modulation of magnetization by gate voltage \cite{ZayetsArch2024_SO_VCMA,MMM2022_SO_VCMA} leads to a negative slope between the gate-voltage-induced changes in $H^0_{ani}$  and $k_{so}$.

It is unlikely that the two aforementioned effects are responsible for the negative slope observed in Fig. \ref{fig:FigDistribution}. The reversal of magnetization alone does not alter the absolute values of either the demagnetization field or the magnetization. Although the exact additional mechanism that changes the anisotropy field with magnetization reversal is not yet entirely clear, it is evident that such a mechanism exists. Furthermore, this mechanism must be sufficiently strong to counteract the opposite contribution to the anisotropy change caused by the variation in the strength of the spin-orbit interaction with magnetization reversal.

The magnetic anisotropy in the studied nanomagnets is primarily created by atoms at the interface of the nanomagnet \cite{Johnson1996Hani,Ohno2010PMA_Thickness}. For magnetic anisotropy to occur, the strength of the spin-orbit interaction must differ significantly between in-plane and perpendicular-to-plane magnetization directions. In the bulk of the amorphous FeCoB nanomagnets, each atom is surrounded symmetrically from all directions, resulting in a symmetric orbital distribution of the bulk atoms. This symmetry leads to a directionally independent spin-orbit interaction strength, meaning these bulk atoms do not contribute to magnetic anisotropy. However, the situation is different for interfacial atoms. The different atoms situated above and below the interface cause the orbitals of interfacial atoms to be asymmetric with respect to the interface normal, creating the necessary anisotropy in the spin-orbit interaction.

An essential condition \cite{LandauField,Zayets2024SObasic} for the existence of spin-orbit interaction is the breaking of time-reversal symmetry (T-symmetry). In the case of the nanomagnets, T-symmetry is broken by the total magnetic field, which is the sum of the internal and external magnetic fields. This condition establishes the important rule that the strength of the spin-orbit interaction is proportional to the degree of T-symmetry breaking and, consequently, to the magnitude of the total magnetic field. This property has been verified both experimentally and theoretically \cite{Zayets2024SObasic} and was applied in the present measurements. The results of this study indicate that the strength of the spin-orbit interaction—and thus the degree of T-symmetry breaking for interfacial atoms—depends not only on the magnitude but also on the polarity of the total magnetic field.

This observed feature is unexpected and intriguing because it is caused by an unusual symmetry breaking in the interfacial atoms. The broken T-symmetry for an orbital means there is a small difference in the distribution of orbital components corresponding to clockwise and counterclockwise electron rotation around the magnetic field. Specifically, one component becomes closer to the atomic nucleus, experiencing a larger electric field from the nucleus and, as a result, a larger magnetic field $H_{so}$ due to spin-orbit interaction than the second component. Since the electron movement is opposite for the two components, each component experiences $H_{so}$ in opposite directions. In the absence of a magnetic field, $H_{so}$ from the two components cancel each other, resulting in no net spin-orbit interaction. Only when the magnetic field breaks T-symmetry does this balance break, causing the electron to experience spin-orbit interaction \cite{Zayets2024SObasic}.

It might be suggested that under the reversal of the magnetic field's polarity, the clockwise and counterclockwise orbital components switch roles. This would mean that the distribution of the counterclockwise component would become identical to the clockwise component before the field reversal. However, our results indicate that this suggestion is incorrect and that there is some non-reciprocity in the breaking of T-symmetry. The distribution of the counterclockwise component before magnetization reversal differs from that of the clockwise component after the magnetization reversal.

Further experimental and theoretical studies are required to understand the details of how such symmetry breaking occurs and to identify the mentioned additional effect that influences anisotropy and varies with magnetization reversal.

In conclusion, we have demonstrated that the strength of the interfacial spin-orbit interaction depends on the direction of magnetization relative to the polarity of the interface. This observation indicates a fundamental difference in interaction strength based on whether the magnetic field penetrates the interface from a ferromagnetic to a non-magnetic metal or vice versa. This effect leads to a variation in magnetic anisotropy for two opposite equilibrium magnetization directions, which we confirmed and measured experimentally.

Our systematic measurements of FeCoB nanomagnets revealed a consistent pattern between changes in magnetic anisotropy and spin-orbit interaction with magnetization reversal. These changes align along a single straight line with a negative slope, suggesting a complex, non-linear relationship. This indicates the existence of an additional, unidentified effect that contributes to the change in magnetic anisotropy with magnetization reversal, beyond the variations in spin-orbit interaction strength.

These findings collectively advance the understanding of spin-orbit interaction and magnetic anisotropy in nanomagnets, opening new avenues for research and potential applications in data storage and spintronic devices.

\bibliography{InterfaceSpinOrbitBib}% Produces the bibliography via BibTeX.

\end{document}